\documentclass[a4paper,12pt]{article}
\usepackage{mathtools}
\usepackage{amssymb}
\usepackage{amsfonts}
\usepackage[footnotesize]{caption}
\usepackage[font=scriptsize]{subcaption}
\usepackage{color}
\usepackage{braket}
\usepackage{cite}
\usepackage{hyperref}
\usepackage{url}
\usepackage{multirow}
\usepackage{relsize}
\usepackage{fullpage}
\usepackage{makecell}
\usepackage{fullpage}

\newcommand{\R}{\mbox{\tiny$R$}}
\newcommand{\T}{\mbox{\tiny$T$}}
\newcommand{\s}{\mbox{\tiny$S$}}
\newcommand{\D}{\mbox{\tiny$D$}}

\begin{document}

\begin{titlepage}
\begin{center}
{\bf\Large $A_4$ Flavour Model for Dirac Neutrinos: Type I and Inverse Seesaw} \\[12mm]
Debasish~Borah$^{\dagger}$
\footnote{E-mail: \texttt{dborah@iitg.ernet.in}},
Biswajit~Karmakar$^{\ddag}$
\footnote{E-mail: \texttt{biswajit@prl.res.in}},
\\[-2mm]

\end{center}
\vspace*{0.50cm}
\centerline{$^{\dagger}$ \it
Department of Physics, Indian Institute of Technology Guwahati,}
\centerline{\it
Assam-781039, India }
\centerline{$^{\ddag}$ \it
Theoretical Physics Division, Physical Research Laboratory,}
\centerline{\it
 Ahmedabad-380009, 
India}

\begin{abstract}
{\noindent
We propose two different seesaw models namely, type I and inverse seesaw to
realise light Dirac neutrinos within the framework of $A_4$ discrete flavour 
symmetry. The additional fields and their transformations under the flavour 
symmetries are chosen in such a way that naturally predicts the hierarchies of 
different elements of the seesaw mass matrices in these two types of seesaw 
mechanisms. For generic choices of flavon alignments, both the models predict 
normal hierarchical light neutrino masses with the atmospheric mixing angle in 
the lower octant. Apart from predicting interesting correlations between 
different neutrino parameters as well as between neutrino and model parameters, 
the model also predicts the leptonic Dirac CP phase to lie in a specific range 
-$\pi/3$ to $\pi/3$. While the type I seesaw model predicts smaller values of 
absolute neutrino mass, the inverse seesaw predictions for the absolute neutrino 
masses can saturate the cosmological upper bound on sum of absolute neutrino 
masses for certain choices of model parameters.
}
\end{abstract}
\end{titlepage}

\section{Introduction}
Although the observations of non-zero neutrino mass and large leptonic mixing
have been confirmed by several neutrino experiments in the last two decades 
\cite{PDG, kamland08, T2K, chooz, daya, reno, minos}, three important issues 
related to neutrino physics are yet not settled. They are namely, (a) nature of 
neutrinos: Dirac or Majorana, (b) mass hierarchy of neutrinos: normal $(m_3 > 
m_2 > m_1)$ or inverted $(m_2 > m_1 > m_3)$ and (c) leptonic CP violation. The 
present status of different neutrino parameters can be found in the latest 
global fit analysis \cite{schwetz16, valle17}. While neutrino oscillation experiments are insensitive to the 
nature of neutrinos, experiments looking for lepton number violating signatures 
can probe the Majorana nature of neutrinos. Neutrinoless double beta decay 
$(0\nu\beta\beta)$ is one such lepton number violating process which has been 
searched for at several experiments without any positive result so far but 
giving stricter bounds on the effective neutrino mass. Cosmology experiments 
are also giving tight constraints on the lightest neutrino mass from the 
measurement 
of the sum of absolute neutrino masses $\sum  m_i  \leq 0.17$ eV 
\cite{Planck15}, disfavouring the quasi-degenerate regime of light neutrino 
masses.

Although negative results at $0\nu \beta \beta$ experiments do not prove that the light neutrinos are of Dirac 
nature, it is nevertheless suggestive enough to come up with scenarios 
predicting Dirac neutrinos with correct mass and mixing. There have been several 
proposals already that can generate tiny Dirac neutrino masses \cite{babuhe, 
diracmass, diracmass1, ma1, ma2, ma3, ma4, db1, db2, db3, db4, 
CentellesChulia:2017koy, type2dirac, A4dirac}. While most of these scenarios 
explain the origin of tiny Dirac mass through some type of seesaw mechanisms at 
tree or loop level, there are some scenarios \cite{diracmass1, A4dirac} which 
consider an additional scalar doublet apart from the standard model (SM) one  
which acquire a tiny vacuum expectation value (vev) naturally due to the 
presence of a softly broken global symmetry. These Dirac neutrino mass models 
also incorporate additional symmetries like $U(1)_{B-L}, Z_N, A_4$ in order to 
generate a tiny neutrino mass 
of purely Dirac type with specific mixing patterns. These symmetries play a
crucial role either in forbidding a tree level Dirac mass term between left 
handed lepton doublet and right handed 
neutrino singlet or a Majorana mass term of right handed neutrino singlet. In 
this work, we particularly look at the possibility of a flavour symmetric 
scenario for Dirac neutrinos within the well motivated $A_4$ flavour symmetry 
group. The details of this non-abelian discrete group is given in Appendix 
\ref{appen1} and can also be found in several review articles 
\cite{discreteRev}. Although there are many $A_4$ realisations of seesaw 
mechanisms for Majorana neutrinos (see~\cite{Karmakar:2014dva} and 
references there in), there are not many studies done in the 
context of Dirac neutrinos. Recently there have been some attempts in this 
direction, specially for type I seesaw \cite{CentellesChulia:2017koy}, type II 
seesaw \cite{type2dirac} and neutrinophilic two Higgs doublet model 
\cite{A4dirac} for Dirac neutrinos.

In the present work, we propose two different seesaw scenarios for Dirac
neutrinos namely, type I and inverse seesaw within the framework of $A_4$ 
flavour symmetry. Type I seesaw for Dirac neutrinos with $A_4$ flavour symmetry 
was also proposed recently by the authors of \cite{CentellesChulia:2017koy} 
along with its correlation to dark matter stability. In this work, we propose a 
more minimal version of type I seesaw as we do not incorporate dark matter into 
account. We also incorporate additional $Z_N$ discrete symmetries in such a way 
that naturally explains the hierarchy of different terms in the neutrino mass 
matrix. Here we note that type I seesaw for Majorana neutrinos were proposed 
long back \cite{ti}. We then propose an inverse seesaw realisation of Dirac 
neutrinos within the $A_4$ flavour symmetric framework. For earlier works on this seesaw 
mechanism for Majorana neutrinos, one may refer to \cite{inverse}. Unlike 
canonical seesaw models, the inverse seesaw can be a low scale framework where 
the singlet heavy neutrinos can be at or below the TeV scale without any fine 
tuning of Yukawa couplings. In the Majorana neutrino scenario, this is possible 
due to softly broken global lepton number symmetry by the singlet mass term. In 
the present case, we however, have a conserved lepton number global symmetry
due to the purely Dirac nature of light neutrinos. Therefore, it is no longer 
possible to use soft $U(1)_L$ global symmetry breaking argument to generate a 
tiny singlet mass term. In spite of that, we generate a tiny singlet neutrino 
mass term at next to leading order by appropriately choosing $Z_N$ discrete 
symmetries. Such discrete symmetries make sure that such a term do not arise at 
leading order so that its smallness can be naturally explained from higher order 
terms.
Similar to the type I seesaw case, here also we can naturally explain the  
hierarchy of different terms present in the inverse seesaw mass matrix. In both 
of these models, the 
antisymmetric term arising out of the products of two $A_4$ triplets plays a 
non-trivial role in generating the correct neutrino mixing. We can obtain the 
tribimaximal (TBM) mixing from the symmetric contribution of the product of the
two triplet flavons while nonzero $\theta_{13}$ is generated from the 
anti-symmetric contribution \cite{A4dirac}. Such anti-symmetric contribution 
from $A_4$ triplet products can play a non-trivial role in generating nonzero 
$\theta_{13}$ in Majorana neutrino scenarios (through Dirac Yukawa 
coupling appearing in type I seesaw) as well \cite{A4Majorana}. The Dirac 
neutrino mass matrix can completely dictate the observed neutrino mixing in this 
construction in case where the charged lepton mass matrix is diagonal. However, 
in some cases, the charged lepton mass matrix can be non-trivial and has an 
important contribution to lepton mixing.

Both of the discrete flavour symmetric constructions for type I and inverse 
seesaw mechanisms show highly predictive nature of the models for generic 
choices of flavon alignments. The anti-symmetric contribution arising from the 
Dirac nature of neutrinos not only generate nonzero $\theta_{13}$ but also shows 
deviations from maximal value of atmospheric mixing angle, favoured by the 
latest global fit data \cite{schwetz16, valle17}. Interestingly, $\theta_{23}$ 
is found to be in the lower octant in our models. Now, due to the particular 
flavour structures of the models, only normal hierarchy for neutrino mass 
spectrum is allowed, another interesting prediction of the model. In addition to 
this, we also constrain the absolute neutrino masses and Dirac CP phase, that 
can be probed at ongoing and future experiments. The model can also be falsified 
by any future observation of $0\nu \beta \beta$.

This letter is organised as follows. In Section \ref{sec:models}, we present 
complete $A_4$ flavour symmetric models for type-I and inverse seesaw scenario 
respectively. Complete phenomenology of the associated models and their 
predictions are also presented in this section. Then we conclude in  Section 
\ref{sec:conc} and included a short note on $A_4$ multiplication rules involved 
in our analysis in the Appendix \ref{appen1}.

\section{$A_4$ Flavour Model with Dirac 
Neutrinos}\label{sec:models}

\subsection{Dirac Type I seesaw}

Unlike in the canonical seesaw mechanism for Majorana neutrinos \cite{ti} where
we incorporate the presence of three (at least two) Majorana heavy neutrinos, 
here we introduce two copies of Weyl fermions $N_{L}$ and $N_{R}$ per 
generation, which are charged under discrete $Z_4 \times Z_3$ symmetry as given 
in Table \ref{tab:t22}. Here $N_{L, R}$ can also be considered to be part of a 
heavy Dirac fermion whose mass can arise either as a bare mass term or from 
flavons depending upon their transformations under the flavour symmetries. In 
Table \ref{tab:t22}, we also show the relevant SM fields, required flavon fields 
as well as their transformations under the flavour symmetry. It can be seen from 
the symmetry transformations that a Dirac mass term for light neutrinos can not 
be written at tree level. However, we can write down mass term for heavy 
neutrinos as well as coupling between light and heavy neutrinos, so that the 
effective light neutrino Dirac mass can be generated from a seesaw mechanism.
\begin{table}[h]
\centering
\resizebox{12cm}{!}{%
\begin{tabular}{|c|cccccc|cccc|}
\hline
 Fields & $L$  & $e_{\R}, \mu_{\R}, \tau_{\R}$ &  $H$ & $\nu_{\R}$& $N_L$ & 
$N_{\R}$ & $\phi_{\s}$ 
& $\phi_{\T}$ &  $\xi$ &$\chi$\\
\hline
$A_{4}$ & 3 & 1,$1''$,$1'$ & 1 & 3 &3 &3& 3 & 3 & 1 &1\\
\hline
$Z_{4}$ & $i$ &-$i$& 1& $1$ & -1& -1&$1$&$1$ &1&-$i$  \\
\hline
$Z_3$ & $\omega$ & $\omega$ & 1& $\omega^2$ & $\omega^2$ & $\omega$& 
$\omega$& 1 & $\omega$ & 1 \\
\hline
\end{tabular}
}\
\caption{\label{tab:t22} Field content and transformation properties under
$A_4 \times Z_4 \times Z_3$ symmetry. }
\end{table}

The relevant Lagrangian for charged lepton sector can be written as 
\begin{equation}\label{Lag:cl2}
 \mathcal{L}_l =  \frac{y_e}{\Lambda}(\bar{L}\phi_{\T})H e_{\R}
+\frac{y_{\mu}}{\Lambda}(\bar{L}\phi_{\T})_{1'}H\mu_{\R}+ 
\frac{y_{\tau}}{\Lambda}(\bar{L}\phi_{\T})_{1''}H\tau_{\R}.
\end{equation}
For generic flavon vev alignment $\langle \phi_T 
\rangle=(v_T,v_T,v_T)$ the corresponding mass matrix is given by 
\begin{eqnarray}\label{mCL2}
m_{l} =\frac{vv_{\T}}{\Lambda} \left(
\begin{array}{ccc}
         y_e & y_{\mu}          & y_{\tau}\\
         y_e & \omega y_{\mu}   & \omega^2 y_{\tau} \\
         y_e & \omega^2 y_{\mu} & \omega y_{\tau} 
\end{array}
\right),
\end{eqnarray}
where $\Lambda$ is the cut-off scale of the theory and $y_e, y_{\mu}, 
y_{\tau}$ are respective coupling constants.
This matrix can be diagonalised by using the magic matrix $U_{\omega}$, given by
\begin{eqnarray}\label{eq:omega}
U_{\omega} =\frac{1}{\sqrt{3}}\left(
\begin{array}{ccc}
         1 & 1          & 1\\
         1 & \omega     & \omega^2\\
         1 & \omega^2   & \omega 
\end{array}
\right).  
\end{eqnarray}
Now, the Lagrangian for neutrino sector can be written as
\begin{eqnarray}
 \mathcal{L}=y_{\D}\chi\bar{L}\tilde{H}N_R /\Lambda + y_{\D'} \chi^2 \bar{N_L} 
\nu_{\R}/\Lambda +y_\xi \xi \bar{N_L}N_{R}+ y_s \phi_{\s} (\bar{N_L}N_{R})_{3s}
 + y_a \phi_{\s} (\bar{N_L}N_{R})_{3a}+\text{h.c.}
\end{eqnarray}
where the subscripts $3s, 3a$ correspond to symmetric and anti-symmetric parts
of triplet products in the $S$ diagonal $A_4$ basis, given in Appendix 
\ref{appen1}. From these contributions, we obtain the mass matrices in $(\nu_L, 
N_R), (N_L, \nu_R), (N_L, N_R)$ basis as  
\begin{eqnarray}\label{xxx}
M_{D}=\frac{y_{\D} vv_{\chi}}{\Lambda}\left(
\begin{array}{ccc}
 1    & 0   & 0 \\
 0       & 1 & 0 \\
 0   & 0   & 1\\
\end{array}
\right), M'_{D}= \frac{y_{\D '} v_{\chi}^2}{\Lambda}\left(
\begin{array}{ccc}
 1    & 0   & 0 \\
 0       & 1 & 0 \\
 0   & 0   & 1\\
\end{array}
\right)~~ {\rm and}
\end{eqnarray}
\begin{eqnarray}\label{xx}
M&=&\left(
\begin{array}{ccc}
 x & 0   & s+a \\
 0   & x & 0\\
 s - a     & 0 & x
\end{array}
\right)~~{\rm with}~~\langle \chi 
\rangle=v_{\chi},\langle \xi 
\rangle=v_{\xi}, ~\langle \phi_S 
\rangle=(0,v_S,0)
\end{eqnarray}
respectively. Such vev alignment for one of the $A_4$ triplet, $\phi_S$ in 
this case, is widely used in the $S$ diagonal basis of $A_4$ and can be realised 
in a natural way by minimisation the scalar potential~\cite{He:2006dk, Branco:2009by, 
A4dirac, Lin:2008aj, Rodejohann:2015hka}. In the $T$ diagonal basis other 
possible vev alignment (e.g. where the first component of the triplet gets vev) 
is adopted~\cite{Altarelli:2005yx}. Here it is worth mentioning that in 
the present set-up  other possible vev alignments like $\langle \phi_S 
\rangle=(v_S,0,0)$ or $\langle \phi_S \rangle=(0,0,v_S)$ are unable to 
reproduce correct neutrino mixing as observed by the experiments. The vev of 
the SM Higgs is denoted by $v$. Here we have defined
$x=y_\xi v_{\xi}$, $s=y_s v_{\s}$, $a=y_a v_{\s}$ and $y_{\D}$ $y_{\D'}$, 
$y_{\xi}$, $y_{s}$, $y_a$ are respective coupling constants involved in the 
neutrino Lagrangian.  Note that $s$ and $a$ are the symmetric and 
antisymmetric contributions originated from $A_4$ multiplication, mentioned earlier. 
This antisymmetric part only contribute in the mass matrix if neutrinos are 
Dirac particles \cite{A4dirac} or in the Dirac neutrino mass matrix used in 
canonical seesaw mechanism for Majorana light neutrinos \cite{A4Majorana}. On 
the other hand, only the symmetric part contributes in a Majorana neutrino mass 
matrix as the anti-symmetric part identically vanishes. Here we will find that 
this antisymmetric part, 
originated due to the Dirac nature of neutrinos, plays an instrumental role in 
the rest of the analysis and crucially dictates the neutrino masses and mixing.
Now, the light Dirac neutrino mass matrix in this type I seesaw like scenario can be 
written as 
\begin{eqnarray}
 m_{\nu}&=&-M'_{D}M^{-1}M_{D}\\
        &=& -\frac{y_{\D}y_{\D'}vv_{\chi}^3}{\Lambda^2}M^{-1}\\
        &=&-\lambda \left(
\begin{array}{ccc}
 x & 0   & -(a+s) \\
 0   & \frac{a^2-s^2+x^2}{x} & 0\\
 a-s     & 0 & x
\end{array}
\right),
\end{eqnarray}
where $\lambda=\frac{y_{\D}y_{\D'}vv_{\chi}^3}{\Lambda^2(a^2-s^2+x^2)}$  is a 
dimensionless quantity. It should be noted that the simple type I seesaw  
formula written above for light Dirac neutrinos is obtained under the assumption 
$M_D, M'_{D} \ll M$ which is justified as the latter is generated at leading 
order whereas $M_D, M'_{D}$ arise at dimension five level only due to the chosen 
particle content and their symmetry transformations. Now we define a Hermitian 
matrix as 
\begin{align}
 \mathcal{M}&=m_{\nu}m_{\nu}^{\dagger}\\
        &=|\lambda|^2 \left(
\begin{array}{ccc}
 |x|^2+|s+a|^2 & 0   & x(a-s)^*-x^*(a+s) \\
 0   & \frac{a^2-s^2+x^2}{x}\frac{(a^2-s^2+x^2)^*}{x^*}& 0\\
 x^*(a-s)-x(a+s)^*& 0 & |x|^2+|a-s|^2
\end{array}
\right).
\end{align}
This matrix can be diagonalised by a unitary matrix $U_{13}$, given by 
\begin{eqnarray}\label{u13}
U_{13}=\left(
\begin{array}{ccc}
 \cos\theta               & 0 & \sin\theta{e^{-i\psi}} \\
     0                    & 1 &            0 \\
 -\sin\theta{e^{i\psi}} & 0 &        \cos\theta
\end{array}
\right),  
\end{eqnarray}
through the relation $U_{13}^{\dagger}\mathcal{M}U_{13}={\rm diag} 
(m_1^2,m_2^2,m_3^2)$. Here we find the mass eigenvalues ($m_1^2,m_2^2, m_3^2$) 
to be
\begin{eqnarray}
m_1^2&=&\kappa^2\left[1+\alpha^2+\beta^2-\sqrt{(2\alpha\beta\cos(\phi_{
ax}-\phi_{ sx}))^2+4(\alpha^2\sin^2\phi_{ax}+\beta^2\cos^2\phi_{sx}) }
\right],\label{eq:tm1}\\
 m_2^2&=&\kappa^2\left[1+\alpha^4+\beta^4+2\alpha^2\cos 2 \phi_{ax} 
 -2\beta^2\cos 2 \phi_{sx}-2\alpha^2\beta^2\cos 2 (\phi_{sx} - \phi_{ax} 
)\right],\label{eq:tm2}\\
 m_3^2&=&\kappa^2\left[1+\alpha^2+\beta^2+\sqrt{(2\alpha\beta\cos(\phi_{
ax}-\phi_{ sx}))^2+4(\alpha^2\sin^2\phi_{ax}+\beta^2\cos^2\phi_{sx}) }
\right],\label{eq:tm3}
\end{eqnarray}
where we have defined $\kappa^2=|\lambda|^2|x|^2$, $\alpha=|a|/|x|$, 
$\beta=|s|/|x|$, $\phi_{sx}=\phi_s - 
\phi_x $, $\phi_{ax}=\phi_a - \phi_x$ with  $s=|s|e^{i\phi_s}$, 
$a=|a|e^{i\phi_a}$ and $x=|x|e^{i\phi_x}$ respectively. From these definitions 
it is clear that  $\alpha$ is associated with the antisymmetric 
contribution whereas $\beta$ is related to the symmetric contribution in the 
Dirac neutrino mass matrix. Now, we obtain the rotation angle and phase involved in $U_{13}$ as 
\begin{eqnarray}\label{eq:th2}
\tan2\theta=\frac{\beta\cos\phi_{sx}\cos\psi-\alpha\sin\phi_{ax}\sin\psi}
        {\alpha\beta\cos(\phi_{sx}-\phi_{ax}) }
\end{eqnarray}
 and 
\begin{eqnarray}\label{eq:tanpsit1}
\tan\psi=-\frac{\alpha\sin\phi_{ax}}
 {\beta\cos\phi_{sx}}
\end{eqnarray}
respectively. 
Now the final lepton mixing matrix is given by 
\begin{eqnarray}
 U&=&U^{\dagger}_{\omega}U_{13}, 
\end{eqnarray}
and the $U_{e3}$ element of the Pontecorvo Maki Nakagawa Sakata (PMNS) leptonic 
mixing matrix is given by $\frac{1}{\sqrt{3}}(\cos\theta+\sin\theta 
e^{-i\psi})$. The PMNS mixing matrix is parametrised as
\begin{equation}
U_{\text{PMNS}}=\left(\begin{array}{ccc}
c_{12}c_{13}& s_{12}c_{13}& s_{13}e^{-i\delta}\\
-s_{12}c_{23}-c_{12}s_{23}s_{13}e^{i\delta}& c_{12}c_{23}-s_{12}s_{23}s_{13}e^{i\delta} & s_{23}c_{13} \\
s_{12}s_{23}-c_{12}c_{23}s_{13}e^{i\delta} & -c_{12}s_{23}-s_{12}c_{23}s_{13}e^{i\delta}& c_{23}c_{13}
\end{array}\right). 
\label{PMNS}
\end{equation}
Comparing $U_{e3}$ from the model with the one in the standard PMNS leptonic mixing matrix $U_{\text{PMNS}}$, we obtain
\begin{eqnarray}
 \sin\theta_{13} e^{-i\delta}=\frac{1}{\sqrt{3}}(\cos\theta+\sin\theta 
e^{-i\psi}).
\end{eqnarray}
Now, $\sin\theta_{13}$ and $\delta$ can be parametrised in terms of $\theta$  
and $\psi$ as 
\begin{eqnarray}\label{eq:s13}
 \sin^2\theta_{13}=\frac{1}{3}(1+\sin 2\theta\cos\psi)~~{\rm and}~~
 \tan\delta=\frac{\sin\theta\sin\psi}{\cos\theta+\sin\theta\cos\psi}.
\end{eqnarray}
\begin{figure}[h]
$$
\includegraphics[height=5cm]{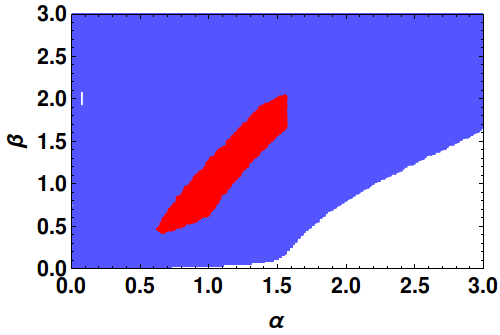}~~~~~~
\includegraphics[height=4.8cm]{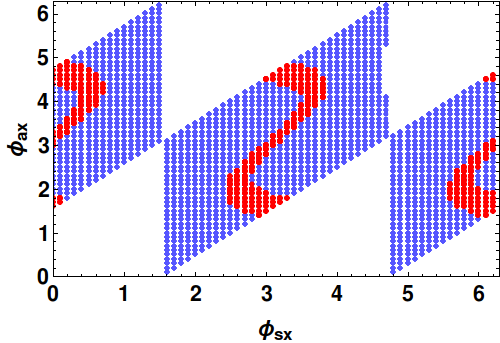}
$$
\caption{Allowed regions of $\beta$-$\alpha$ (left panel) and  
$\phi_{ax}$-$\phi_{sx}$ (right panel) planes from the 3$\sigma$ global fit 
values of $\theta_{13}$, 
$\theta_{12}$ and $\theta_{23}$~\cite{valle17} represented by the blue dots. 
Red dots in each plot also satisfy $3\sigma$ allowed range for the ratio 
($r$) of solar to atmospheric mass squared differences~\cite{valle17}.}
\label{fig:abpp}
\end{figure}
Such correlation between $\sin\theta_{13}$ 
($i.e.$ $U_{e3}$) and the model parameters can easily be obtained and can also be found 
in~\cite{A4dirac, Grimus:2008tt,Albright:2008rp,Albright:2010ap,He:2011gb}
for other scenarios. From equations (\ref{eq:th2}-\ref{eq:s13}) it is clear 
that, all 
the mixing angles ($\theta_{13}, \theta_{12}, \theta_{23}$) and Dirac CP phase 
$(\delta)$ involved in the lepton mixing matrix $U_{\text{PMNS}}$ are 
functions of four parameters namely, $\alpha$, $\beta$, $\phi_{ax}$ and 
$\phi_{sx}$. Now, using 3$\sigma$ allowed range~\cite{valle17} of the three 
mixing angles ($\theta_{13}, \theta_{12}, \theta_{23}$), in figure \ref{fig:abpp} 
we have shown the constrained range of $\alpha$, $\beta$, $\phi_{ax}$ and
$\phi_{sx}$. In figure \ref{fig:abpp}, the blue dots represent
the allowed points in the $\alpha$-$\beta$ plane (left panel) and 
$\phi_{ax}$-$\phi_{sx}$ (right panel)
plane respectively. In addition to the bounds obtained from the mixing angles, 
the parameter space can be further constrained in order to satisfy the ratio of 
solar to atmospheric mass squared differences, defined as 
\begin{eqnarray}\label{eq:r}
r=\frac{\Delta{m}_{\odot}^{2}}{|\Delta{m}_{A}^{2}|}= 
\frac{\Delta{m_{21}^{2}}}{|\Delta{m}^2_{31}|}. 
\end{eqnarray}
From equation (\ref{eq:tm1}-\ref{eq:tm3}) and equation (\ref{eq:r}) it is evident that 
this ratio $r$ is also function of $\alpha$, $\beta$, $\phi_{sx}$ and $\phi_{ax}$ (which are 
appearing in the expression for the mixing angles). Once again using the 
\begin{figure}[h]
$$
\includegraphics[height=5.0cm]{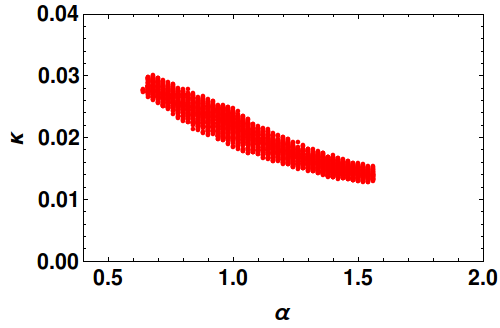}~~~~~
\includegraphics[height=4.8cm]{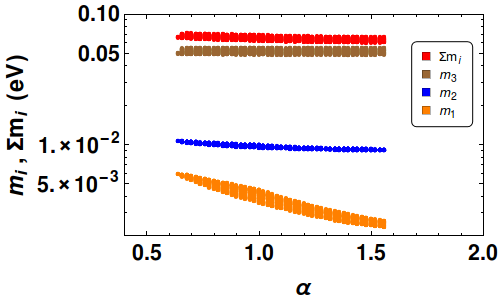}
$$
\caption{Left panel: Estimation for $\kappa$ (in eV) as a function 
of $\alpha$. Right panel: Prediction for absolute neutrino masses (orange, blue, 
brown and red for $m_1$, $m_2$, $m_3$ and $\sum m_i$ 
respectively) as a function of $\alpha$. In both cases the parameter space 
simultaneously satisfies 3$\sigma$ allowed range of $\theta_{13}$, 
$\theta_{12}$, $\theta_{23}$ and $r$~\cite{valle17} as shown in figure 
\ref{fig:abpp}.}
\label{fig:k-m-t1}
\end{figure}
$3\sigma$ range of the neutrino mass squared differences we find the 
allowed ranges for $\alpha$, $\beta$, $\phi_{sx}$ and $\phi_{ax}$ given by the 
red dots in the both panels of figure \ref{fig:abpp}. Therefore, these 
red dots represents the regions of model parameters that satisfy the complete 
neutrino oscillation data~\cite{valle17}. This reveals that the allowed range 
of $\alpha \approx$ 0.6-1.6  corresponds to $\beta \approx$ 0.4-2.0
 as evident for the left panel of figure \ref{fig:abpp}. On the 
other hand, the right panel plot
of figure \ref{fig:abpp} shows that few disconnected regions are allowed in the 
$\phi_{sx}$-$\phi_{ax}$ parameter space.

\begin{figure}[h]
$$
\includegraphics[height=5cm]{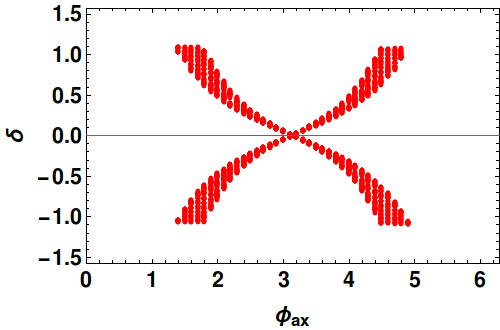}~
\includegraphics[height=5cm]{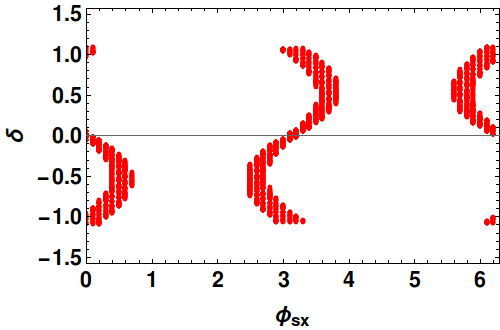}
$$
$$
\includegraphics[height=5cm]{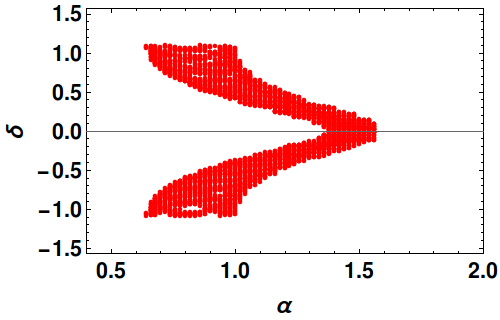}
$$
\caption{Predictions for Dirac CP phase $\delta$ (in radian) as a function of 
$\phi_{ax}$, $\phi_{sx}$ and $\alpha$ for 3$\sigma$ allowed range of 
$\theta_{13}$, $\theta_{12}$, $\theta_{23}$ and $r$~\cite{valle17} as evaluated 
in figure \ref{fig:abpp}.}
\label{fig:delt1}
\end{figure}

Now, using these allowed values (obtained from figure \ref{fig:abpp}) for the 
parameters ($\alpha$, $\beta$, $\phi_{sx}$ and $\phi_{ax}$) and the best fit 
value for the solar mass squared difference $\Delta{m}_{\odot}^{2}= 
7.5\times10^{-5}~ $eV$ ^2$~\cite{valle17}, we can find the the common factor 
$\kappa$ appearing in the absolute 
neutrino mass eigenvalues using the relation
\[
  \kappa=
  \sqrt{
    \begin{aligned}
    \Delta{m}_{\odot}^{2}/&\{\left[1+\alpha^4+\beta^4+2\alpha^2\cos 
2 
\phi_{ax} 
 -2\beta^2\cos 2 \phi_{sx}-2\alpha^2\beta^2\cos 2 (\phi_{sx} - \phi_{ax} 
)\right] \\
     & -[1+\alpha^2+\beta^2-\sqrt{(2\alpha\beta\cos(\phi_{
ax}-\phi_{ sx}))^2+4(\alpha^2\sin^2\phi_{ax}+\beta^2\cos^2\phi_{sx}) }
]\}
    \end{aligned}
  }.
\]
Here we have used equations (\ref{eq:tm1}) and \eqref{eq:tm2} to deduce the above 
correlation. In left panel of figure \ref{fig:k-m-t1} we have plotted the allowed 
values for $\kappa$ (in eV) as a function of $\alpha$, where we also find that 
the allowed range $\alpha\approx$ 0.6-1.6 restricts $\kappa$ to fall in the range 
0.012-0.03 eV. Now using the estimation for $\kappa$ as in left-panel of figure
\ref{fig:k-m-t1}, we can find the absolute neutrino masses using equations 
(\ref{eq:tm1})-\eqref{eq:tm3}. In the right panel of figure \ref{fig:k-m-t1} we 
have plotted the individual absolute neutrino masses (where orange, blue, brown 
dots stand for $m_1$, $m_2$ and $m_3$ respectively) as well as their sum 
($\sum m_i$ denoted by the red dots) as a function of $\alpha$. Here we find 
that the allowed ranges for the absolute neutrino masses (obeying normal 
hierarchy) are given by 
$m_1\approx0.0060-0.0023$ eV, $m_2\approx 0.0105-0.0090$ eV, $m_3\approx 
0.0547-0.0481$ eV and $\sum m_i\approx 0.0707-0.0596$ eV when $\alpha$ is in 
the range 0.6-1.6 . In the present setup, an inverted hierarchy of  
light neutrino mass spectrum however can not be accommodated, an interesting 
prediction that will undergo tests in several ongoing and near future 
experiments.

\begin{figure}[h]
$$
\includegraphics[height=5cm]{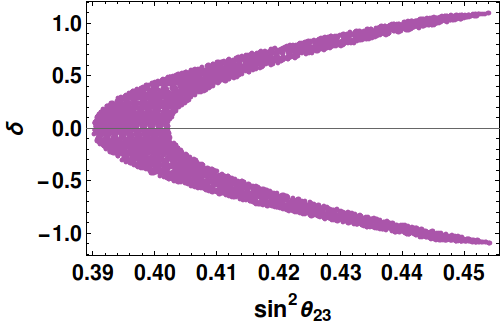}~
\includegraphics[height=5cm]{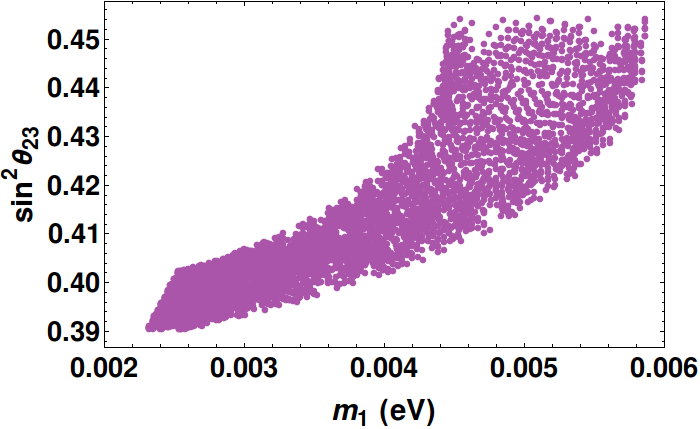}
$$
$$
\includegraphics[height=5cm]{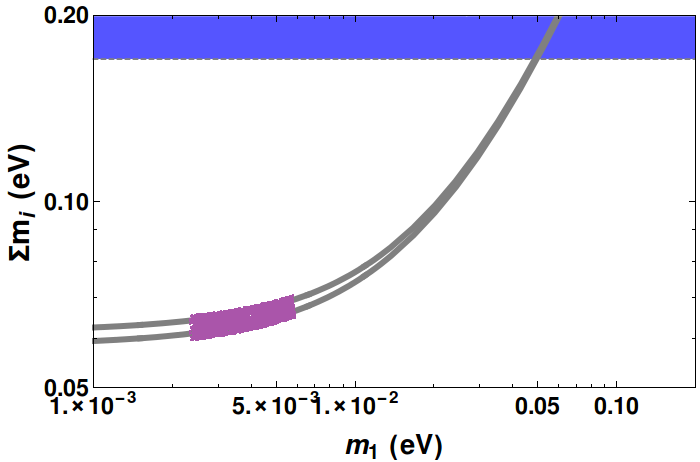}
\includegraphics[height=5cm]{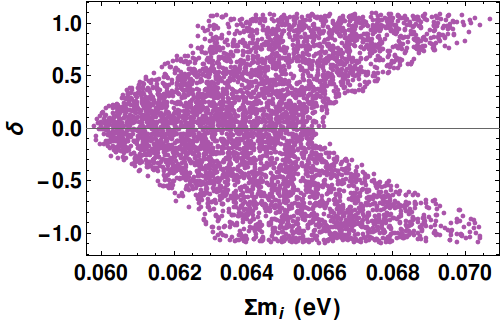}
$$
\caption{Correlations between different light neutrino parameters for 3$\sigma$ allowed range of $\theta_{13}$, 
$\theta_{12}$, $\theta_{23}$ and $r$~\cite{valle17}.}
\label{fig:colt1}
\end{figure} 

Now, from equations (\ref{eq:tanpsit1}) and (\ref{eq:s13}), we find that the Dirac CP 
phase can evaluated once we find allowed parameter space in the present model. 
Therefore using the allowed regions for $\alpha$, $\beta$, $\phi_{sx}$ and 
$\phi_{ax}$ as obtained in figure \ref{fig:abpp}, we can find the predictions for 
the Dirac CP phase $\delta$ within this framework. In figure \ref{fig:delt1}
we have shown the prediction for $\delta$ as a function of $\phi_{ax}$, 
$\phi_{sx}$, $\alpha$ and it is clear that the model predicts $\delta$ to be in the range 
-$\pi/3\lesssim \delta \lesssim \pi/3$.

Next, to understand the correlation between associated observables in the 
present type I seesaw framework we present few schematics in figure 
\ref{fig:colt1}. Here the upper left panel represents the correlation between 
$\delta$ and $\theta_{23}$. This figure shows that in our setup $\theta_{23}$ 
falls in the lower octant when $\delta$ lies between -$\pi/3$ to $\pi/3$.
This is in good agreement with all three global fit analysis~\cite{schwetz16,valle17, 
Capozzi:2016rtj} where the best fit value for $\theta_{23}$ prefers to be in 
the lower octant (although for 3$\sigma$ range both octants are possible) for 
the normal hierarchy of light neutrino masses. And in our case, only normal 
hierarchy of neutrino mass spectrum is allowed. Now, in the other two panels of 
figure \ref{fig:colt1} we have plotted our allowed parameter space in 
$\sin^2\theta_{23}$-$m_1$ and $\delta$-$\sum m_i$ plane respectively. From 
$\sin^2\theta_{23}$ versus $m_1$ plot it is clear that as $m_1$ approaches towards 
its maximum, $\theta_{23}$ also tends towards the maximal value. Now in 
the left panel of figure \ref{fig:colt1} we plot sum of the absolute neutrino 
masses $\sum m_i$ as a function of lightest light neutrino mass $m_1$ and it 
falls well below the Planck upper limit \cite{Planck15} (as shown by the shaded 
region). In this figure the splitting in the sum of absolute mass is due to 
3$\sigma$ uncertainties in the solar and atmospheric mass squared 
differences~\cite{valle17}. Now, on the other hand, $\delta$ versus $\sum m_i$ 
plot in the lower right panel of figure   \ref{fig:colt1} shows that for 
$\delta$ within the range -$\pi/3$ to $\pi/3$,  $\sum m_i$ ranges between 
0.0707 eV to 0.0596 eV indicating higher value of $\sum m_i$ is allowed only 
when $\delta \neq 0$. It is interesting to note that, the predicted values of 
$\sum m_i$ lie well below the cosmological upper bound $\sum  m_i  \leq 0.17$ 
eV \cite{Planck15}.

\subsection{Dirac Inverse seesaw}
In usual inverse seesaw model, the complete neutral fermion mass matrix is
$9\times9$ whose structure in the $(\nu_L, N_R, S_R)$ basis
            \begin{equation}\label{eq:2}
      M_{\nu}= \left(\begin{array}{ccc}
      0 & m^{T}_{D} & 0 \\
      m_{D}& 0 & M^{T}\\ 
      0 & M & \mu
      \end{array}\right)
      \end{equation}
where $m_D$ is the usual Dirac neutrino mass. The lepton number violation 
occurs only through the $3 \times 3$ block denoted by $\mu$ so that this term 
can be naturally small. Block diagonalisation of the above mass matrix results 
in the effective light neutrino mass matrix as ,
       \begin{equation}\label{eq:iss1} 
        m_{\nu} = m_{D}^{T}(M^{T})^{-1} \mu M^{-1}m_{D}
       \end{equation}
Unlike canonical seesaw where the light neutrino mass is inversely proportional 
to the lepton number violating Majorana mass term of singlet neutrinos, here 
the light neutrino mass is directly proportional to the singlet mass term 
$\mu$. The heavy neutrino masses are proportional to $M$. Here, even if $M \sim 
1$ TeV, correct neutrino masses can be generated for $m_D \sim 10$ GeV, say if 
$\mu \sim 1$ keV. Such small $\mu$ term is natural as $\mu \rightarrow 0$ helps 
in recovering the global lepton number symmetry $U(1)_L$ of the model. Thus, 
inverse seesaw is a natural TeV scale seesaw model where the heavy neutrinos 
can remain as light as a TeV and Dirac mass can be as large as the charged 
lepton masses and can still be consistent with sub-eV light neutrino masses. 

In this section, we wish to construct a similar mass matrix for Dirac neutrinos
so that the smallness of light Dirac neutrino mass can be generated naturally 
by a TeV scale seesaw. Since lepton number is conserved for Dirac neutrinos, we 
consider it as a conserved global symmetry of the model, similar to the type I 
seesaw discussed above. The field content of the proposed model is given in 
table \ref{tab:inverse}. The $A_4$ symmetry is augmented by $Z_4 \times Z_3 
\times Z_2$ discrete symmetries in order to make sure that the desired strengths 
of different elements of the inverse seesaw mass matrix are naturally obtained. 

\begin{table}[h]
\centering
\resizebox{14cm}{!}{%
\begin{tabular}{|c|cccccccc|cccccc|}
\hline
 Fields & $L$  & $e_{\R}, \mu_{\R}, \tau_{\R}$ &  $H$ & $\nu_{\R}$& $N_L$ & 
$N_{\R}$ & $S_L$ & $S_R$ & $\phi_{\s}$ & $\phi_{\T}$ &  $\xi$ & $\zeta$ &$\eta$ 
& $\phi^{\prime}$ \\
\hline
$A_{4}$ & 3 & 1,$1''$,$1'$ & 1 & 3 &3 &3& 3 & 3 & 3 & 3 & 1 &1 &1 & 1\\
\hline
$Z_{4}$ & $1$ &1 & 1 & $i$& -$i$  & 1& $1$ & -$i$ & 1 & 1 & 1&1 & -1 & $i$  
\\
\hline
$Z_3$ & $\omega$ & $\omega$ & 1 & 1& 1& 1 & 1 & 1 & $\omega$ & 1& 
$\omega$&1 & 1 & $\omega$ \\
\hline
$Z_2$ & 1 & 1 & 1 & -1& 1& -1 & 1 & -1 & -1 & 1 & -1&-1 & 1 & -1 \\
\hline
$U(1)_L$ & 1 & 1& 0 &1 &1 &1 &1 &1 & 0 & 0 & 0 & 0& 0 & 0 \\
\hline
\end{tabular}
}\
\caption{\label{tab:inverse}  Fields content and transformation properties under
$A_4 \times Z_4 \times Z_3 \times Z_2$ symmetry for inverse seesaw. }
\end{table}

The Lagrangian for the above field content can be written as 
\begin{align}\label{eq:isslag}
 \mathcal{L}_Y &= \frac{y_e}{\Lambda}(\bar{L}\phi_{\T})H e_{\R}
+\frac{y_{\mu}}{\Lambda}(\bar{L}\phi_{\T})_{1'}H\mu_{\R}+ 
\frac{y_{\tau}}{\Lambda}(\bar{L}\phi_{\T})_{1''}H\tau_{\R}+\frac{\bar{L}\tilde{H
}N_R}{\Lambda} 
 \left(y_x\xi+ y_s\phi_S+y_a\phi_S\right) \nonumber \\ 
& + \frac{Y_{RN}}{\Lambda}  \bar{\nu_{\R}}{N_L}\eta \zeta + 
Y_{NS} \bar{S_R} N_L \zeta+Y^{\prime}_{NS} \bar{S_L} N_R \zeta + 
\frac{Y_{S}}{\Lambda^2} \bar{S_L} S_R \phi^{\prime 3}.
\end{align}

We consider the vev alignment (similar to the one present in the 
previous subsection) of the flavons as
\begin{eqnarray}
 \langle \phi_T \rangle= (v_T, v_T, v_T), \langle \phi_S \rangle=(0, 
v_S,0), \langle \xi \rangle=v_{\xi}, \langle \zeta \rangle=v_{\zeta}, 
\langle \eta \rangle=v_{\eta}, \langle \phi' \rangle= v_{\phi'}.
\end{eqnarray}
Effective light neutrino mass in this scenario can be written as,
\begin{equation}\label{eq:nuiss}
        m_{\nu} = M_{RN}(M^{\prime}_{NS})^{-1} M_S M^{-1}_{NS} M_{\nu N}. 
\end{equation}
From the Lagrangian presented in equation (\ref{eq:isslag}), we can find the  
mass matrices involved in the neutrino sector after symmetry breaking ($A_4$ as 
well as electroweak) as
$$ M_{RN} = \frac{Y_{RN}}{\Lambda}  
v_\eta v_S \mathbf{I}, 
M_{NS} = Y_{NS} v_\zeta \mathbf{I}, M^{\prime}_{NS} = Y^{\prime}_{NS} 
v_{\zeta}\mathbf{I}, M_S =  \frac{Y_{S}}{\Lambda^2} 
v_{\phi^{\prime}}^3\mathbf{I}
$$

\begin{eqnarray}\label{eq:MnuN}
 M_{\nu N} = \frac{v v_S}{\Lambda}\left(
\begin{array}{ccc}
 x & 0   & s+a \\
 0  & x & 0\\
 s - a & 0 & x
\end{array}
\right).
\end{eqnarray}
Here, $x=y_\xi v_{\xi}$, $s=y_s v_{\s}$ and $a=y_a 
v_{\s}$ respectively where $s$ and $a$ stands for symmetric and antisymmetric 
contributions originated from $A_4$ multiplication similar to the type I seesaw  
case discussed before. The couplings $Y_{RN}, Y_{NS}, Y^{\prime}_{NS}, Y_{S}, 
y_\xi, y_s, y_a$ are the Yukawa couplings given in the above Lagrangian and 
$\Lambda$ is the cut-off scale. Again, here we emphasise that the antisymmetric 
part of $A_4$ triplet products particularly contribute to any Dirac type mass 
matrix involved in the neutrino seesaw formula and the associated phenomenology 
crucially depends on this contribution. Since the construction of the charged 
lepton sector is exactly identical with type I seesaw scenario, it can again be 
diagonalised by the magic matrix $U_{\omega}$ given in equation 
(\ref{eq:omega}).  To diagonalise the neutrino mass matrix let us define the 
Hermitian mass matrix as before
\begin{eqnarray}\label{eq:mmdiss}
 \mathcal{M}=m_{\nu}m_{\nu}^{\dagger}=|\lambda|^2
 \left(
 \begin{array}{ccc}
 |x|^2+|s + a|^2 & 0   & x(s - a)^*+x^*(s+a) \\
 o& |x|^2 & 0\\
 x^*(s - a)+x(s+a)^*& 0 & |x|^2+|s - a|^2
\end{array}
\right),
\end{eqnarray}
where $\lambda=\frac{Y_{RN} Y_{S}}{Y_{NS}Y'_{NS}}\frac{vf^3}{\Lambda^4}$. Here 
we have assumed the vev of all the scalar flavons (except the SM Higgs) to be same and 
denoted by $f$, $i.e$,   $v_{\s}=v_{\xi}=v_{\zeta}=v_{\eta}=v_{\phi'}=f$. 
The Hermitian matrix $\mathcal{M}$ can be diagonalised by a unitary matrix 
$U_{13}$ as given in equation (\ref{u13}), obeying $U_{13}^{\dagger}\mathcal{M}U_{13}={\rm diag} 
(m_1^2,m_2^2,m_3^2)$, 
where the two parameters $\theta$ and $\psi$ appearing in $U_{13}$ are found to 
be
\begin{eqnarray}\label{eq:angiss}
 \tan 2\theta=\frac{\alpha\sin\phi_{ax}\sin\psi-\beta\cos\phi_{sx}\cos\psi}
 {\alpha\beta\cos(\phi_{sx}-\phi_{ax})}~~~ {\rm and}~~~ 
\tan\psi=-\frac{\alpha\sin\phi_{ax}}
 {\beta\cos\phi_{sx}}.
\end{eqnarray} 
Here,  $\alpha=|a|/|x|$, $\beta=|s|/|x|$, $\phi_{sx}=\phi_s - 
\phi_x $, $\phi_{ax}=\phi_a - \phi_x$ with  $s=|s|e^{i\phi_s}$, 
$a=|a|e^{i\phi_a}$ and $x=|x|e^{i\phi_x}$ respectively. 
Hence  $\alpha$ is basically associated with the antisymmetric 
contribution whereas $\beta$ is related to the symmetric contribution in the 
Dirac neutrino mass matrix. The final lepton 
mixing matrix in this case is also governed by the mixing matrix, 
$U=U^{\dagger}_{\omega}U_{13}$ involving contributions from both charged 
lepton and neutrino sector. Therefore, the correlation of $\theta_{13}$ (and 
$\delta$) with $\theta$ (and $\psi$) in this case is similar to the one 
presented in the type I seesaw case as given by equation \ref{eq:s13}. 

\begin{figure}[h]
$$
\includegraphics[height=5cm]{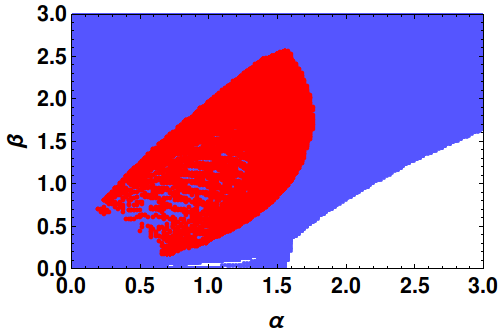}~~~~~~
\includegraphics[height=4.8cm]{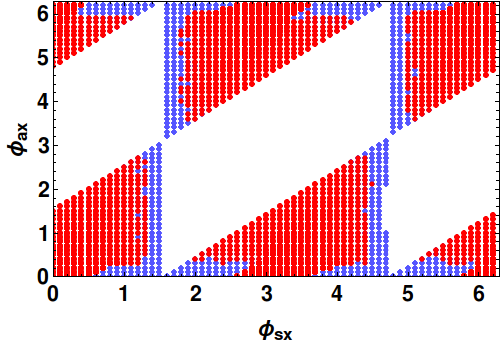}
$$
\caption{Allowed regions of $\beta$ vs $\alpha$ (left panel) and $\phi_{ax}$ vs 
$\phi_{sx}$ (right panel) for 3$\sigma$ allowed range of $\theta_{13}$, 
$\theta_{12}$ and $\theta_{23}$ represented by the blue dots. Red dots in each 
plot also satisfies $3\sigma$ allowed range for the the solar to atmospheric 
mass-squared ratio $r$ along with upper limit on sum of the thee light 
neutrinos $\sum m_i \leq 0.17$ eV~\cite{Planck15}, representing the actual 
allowed parameter space.}
\label{fig:abpiss}
\end{figure}

After diagonalisation of the Hermitian matrix as given in equation 
(\ref{eq:mmdiss}), the real, positive squared mass eigenvalues are obtained as
\begin{eqnarray}
m_1^2&=&\kappa^2\left[1+\alpha^2+\beta^2-\sqrt{(2\alpha\beta\cos(\phi_{
ax}-\phi_{ sx}))^2+4(\alpha^2\sin^2\phi_{ax}+\beta^2\cos^2\phi_{sx}) }
\right],\label{eq:issm1}\\
 m_2^2&=&\kappa^2,\label{eq:issm2}\\
 m_3^2&=&\kappa^2\left[1+\alpha^2+\beta^2+\sqrt{(2\alpha\beta\cos(\phi_{
ax}-\phi_{ sx}))^2+4(\alpha^2\sin^2\phi_{ax}+\beta^2\cos^2\phi_{sx}) }
\right],\label{eq:issm3}
\end{eqnarray}
where we have defined $\kappa^2=|\lambda|^2|x|^2$. Here we find that, 
both neutrino mixing angles and masses are functions of parameters like $\alpha$, 
$\beta$, $\phi_{ax}$ and $\phi_{sx}$ as evident from equations (\ref{eq:angiss}) 
 and (\ref{eq:issm1}-\ref{eq:issm3}) respectively. Using similar strategy, we 
again try to constrain the involved parameter space ($\alpha$, $\beta$, 
$\phi_{ax}$ and $\phi_{sx}$) as illustrated in figure \ref{fig:abpiss}. The blue 
dots in both left (in $\alpha$-$\beta$ plane) and right (in 
$\phi_{sx}$-$\phi_{ax}$ plane) panel satisfies 3$\sigma$ allowed range for the 
neutrino mixing angles~\cite{valle17}. Then we impose the constraints 
(varying within $3\sigma$ range) coming from the ratio of the two mass squared 
differences as defined in equation (\ref{eq:r}). The red dots in both the panels of 
figure \ref{fig:abpiss} shows allowed ranges of the parameter space, after  
taking both these constraints (mixing angles and mass squared difference ratios) 
into account. In the left panel of figure \ref{fig:abpiss} we find that, 
corresponding to $\alpha$ in 
the range 0.2 to 1.7, $\beta$ is restricted within 2.5.  The 
right panel of the same plot reveals that a few disconnected regions in the 
$\phi_{sx}$-$\phi_{ax}$ plane are allowed. Note that here we have also used the
\begin{figure}[h]
$$
\includegraphics[height=5cm]{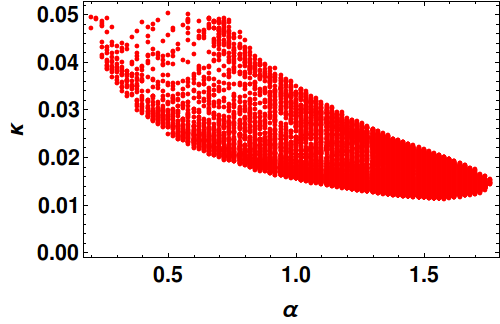}~~~~~
\includegraphics[height=5cm]{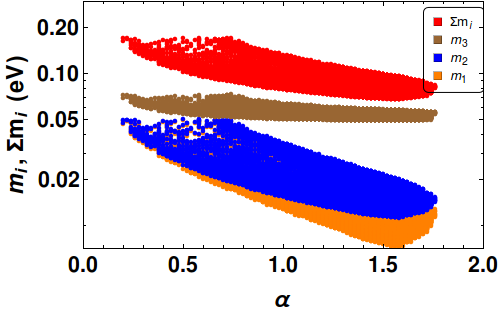}
$$
\caption{Left panel: Estimation for $\kappa$ (in eV) as a function 
of $\alpha$. Right panel: Prediction for absolute neutrino masses ( 
orange, blue, 
brown and red for $m_1$, $m_2$, $m_3$ and $\sum m_i$ respectively.) and Dirac 
CP phase $\delta$ (right panel) for 3$\sigma$ allowed range of $\theta_{13}$, 
$\theta_{12}$, $\theta_{23}$, $r$~\cite{valle17} along with with upper limit on 
sum of the thee light 
neutrinos $\sum m_i \leq 0.17$ eV~\cite{Planck15}
}.
\label{fig:k-m-iss}
\end{figure}
recent upper bound on sum of the thee light neutrinos $\sum m_i \leq 0.17$ 
eV~\cite{Planck15} to constrain the parameter space and afterwards we analyse 
only those regions which satisfy this limit. Now, the common factor 
($\kappa$) appearing in the neutrino mass eigenvalues shown in equations 
(\ref{eq:issm1}-\ref{eq:issm3}) can be evaluated using 
\begin{eqnarray}
 \kappa= 
\sqrt{\Delta{m}_{\odot}^{2}/\{1-[1+\alpha^2+\beta^2-\sqrt{
(2\alpha\beta\cos(\phi_{
ax}-\phi_{ sx}))^2+4(\alpha^2\sin^2\phi_{ax}+\beta^2\cos^2\phi_{sx}) }
]\}}. 
\label{kappa34}
\end{eqnarray}
In figure \ref{fig:k-m-iss} left panel we show the estimates of $\kappa$ (in eV) 
as a function of $\alpha$. Also, it is worth mentioning that due to particular 
flavour structure of the this inverse seesaw scenario $m_2$ coincides with $\kappa$ as 
given in equation \ref{eq:issm2}. Now, our prediction for absolute neutrino masses 
(with orange, blue, brown 
dots representing $m_1$, $m_2$ and  $m_3$ respectively) 
and their sum ($\sum m_i$ denoted by the red dots) are given in the 
right panel of figure \ref{fig:k-m-iss}. Here we find 
that the allowed ranges for the absolute neutrino masses (obeying normal 
hierarchy) are given by 
$m_1\approx 0.050-0.007$ eV, $m_2\approx 0.051-0.010$ eV, $m_3\approx 
 0.072-0.049$ eV and $\sum m_i\approx 0.17-0.067$ eV when $\alpha$ is in 
the range 0.2-1.7. In this inverse seesaw scenario, inverted   mass 
hierarchy is not possible as $\Delta 
 m^2_{23}+\Delta m^2_{21}=-2k^2(\alpha^2+\beta^2)<0$.

\begin{figure}[h]
$$
\includegraphics[height=5cm]{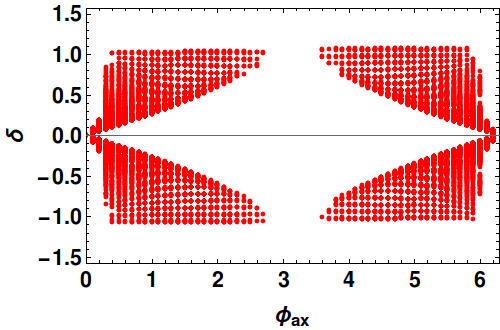}~
\includegraphics[height=5cm]{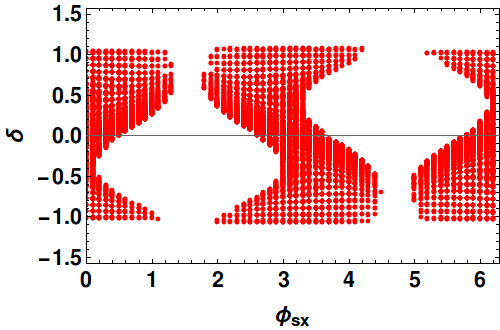}
$$
$$
\includegraphics[height=5cm]{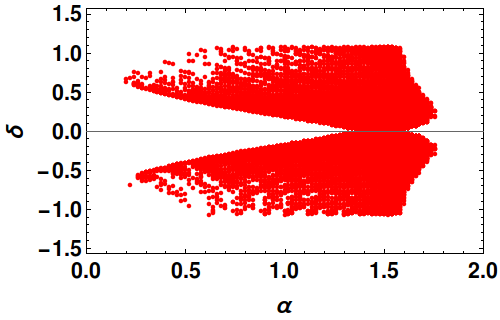}
$$
\caption{Predictions for  Dirac CP phase $\delta$ (in radian) as a function of 
$\phi_{ax}$, $\phi_{sx}$ and $\alpha$ for 3$\sigma$ allowed range of 
$\theta_{13}$, 
$\theta_{12}$, $\theta_{23}$, $r$~\cite{valle17} along with the limit on 
sum of the thee light 
neutrinos $\Sigma m_i\leq 0.17$ eV~\cite{Planck15}.}
\label{fig:deliss}
\end{figure}

Now, to illustrate the prediction for Dirac CP phase and its dependence on 
the parameters of the model, in figure \ref{fig:deliss} we present the allowed 
regions for $\delta$ as a function of $\phi_{ax}$ (upper left panel), 
$\phi_{sx}$ (upper right panel) and $\alpha$ (bottom panel) respectively. From 
these plots it turns out that in this inverse seesaw scenario, the allowed value 
 for $\delta$ lies in appropriately range $-\pi/3$ to $+\pi/3$, similar to what 
we saw for type I seesaw case before. 
\begin{figure}[h]
$$
\includegraphics[height=5cm]{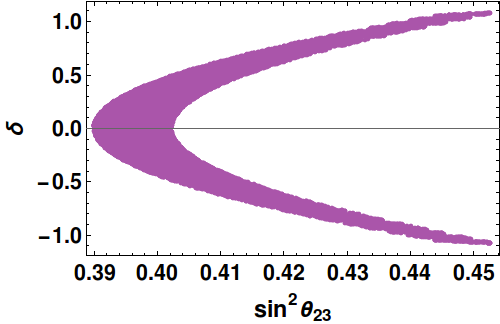}~
\includegraphics[height=5cm]{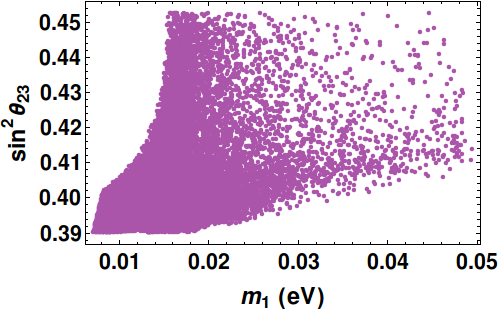}
$$
$$
\includegraphics[height=5cm]{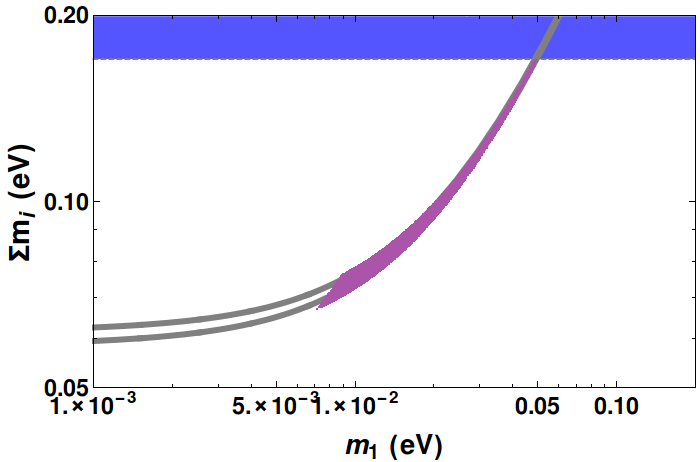}
\includegraphics[height=5cm]{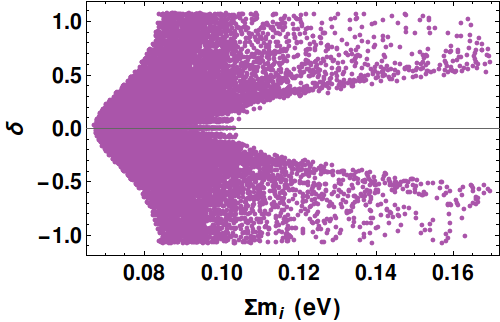}
$$
\caption{Correlations between different light neutrino parameters for 3$\sigma$ 
allowed range of $\theta_{13}$, $\theta_{12}$, $\theta_{23}$, $r$~\cite{valle17} 
along with 
upper limit on sum of the thee light neutrinos $\sum m_i \leq 0.17$ 
eV~\cite{Planck15}.}
\label{fig:coliss}
\end{figure}

Finally, to understand the correlation between observables associated with 
neutrino masses and mixings in this inverse framework, we refer to figure 
\ref{fig:coliss}. In the upper left panel of this figure a correlation between  
$\delta$ and $\theta_{23}$ is presented and we find that for $\delta$ in the  
range -$\pi/3$ to $\pi/3$, $\theta_{23}$ always falls in the lower octant. 
As mentioned earlier, this is in good agreement with all three global 
analysis~\cite{schwetz16, valle17, Capozzi:2016rtj} where the best fit value for 
$\theta_{23}$ prefers to be in the lower octant (although for 3$\sigma$ range 
both the octants are possible) for the normal hierarchy of light neutrino masses. 
Here we remind ourself that only normal hierarchy of neutrino mass is allowed 
in the present scenario. In the upper right panel of figure \ref{fig:coliss},  
we have plotted the allowed parameter space in $\sin^2\theta_{23}$-$m_1$ plane 
whereas the bottom left panel 
represents the allowed region in the $\delta$-$\sum m_i$ plane. From 
$\sin^2\theta_{23}$ versus $m_1$ plot it is clear that smaller the lightest 
neutrino mass $m_1$,  more likely is the deviation of $\theta_{23}$ from its 
maximal value. In the bottom left panel of figure \ref{fig:coliss}, the 
purple dots show the model predictions for $\sum m_i$ corresponding to the 
lightest light neutrino mass $m_1$, representing a high mass regime for the 
light neutrinos. Here, the region bounded by the solid lines represent 
3$\sigma$ 
uncertainty in the mass squared differences and the shaded region stands for the 
disallowed 
region by the Planck upper limit~\cite{Planck15}.
Finally, $\delta$ vs $\sum m_i$ plot in the bottom right panel shows that all 
regions of $\delta$ (between -$\pi/3$ to $\pi/3$) allowed with $\sum m_i$ 
ranging in between  0.067 eV to 0.17 eV indicating higher value of $\sum m_i$ 
is only possible when $\delta \neq 0$. Such high values of $\sum m_i$ can 
saturate the  cosmological upper bound $\sum  m_i  \leq 0.17$ eV 
\cite{Planck15} which can indirectly constrain the Dirac CP phase as well. 

It is observed that the allowed range of the lightest neutrino 
mass is different in inverse seesaw case compared to what is obtained for type I 
seesaw. This is evident from the right panels of figure \ref{fig:k-m-t1} and 
figure \ref{fig:k-m-iss} for type I and inverse seesaw respectively. This can be 
explained from the difference in light neutrino mass eigenvalue expressions 
given in equations \eqref{eq:tm1}, \eqref{eq:tm2}, \eqref{eq:tm3} for type I 
seesaw and equations \eqref{eq:issm1}, \eqref{eq:issm2}, \eqref{eq:issm3} for 
inverse seesaw. As can be seen from these expressions, the second mass 
eigenvalue ($m_2$) expression is very different in the two cases due to the 
$A_4$ flavour symmetric construction and the governing 
seesaw mechanism. Due to this difference, constraint coming from the ratio of 
solar to atmospheric mass squared differences ($r$) in these two scenarios are 
such that the inverse seesaw scenario permits a relatively larger allowed 
parameter space (for $\alpha$ and $\beta$) satisfying neutrino oscillation 
data. This is evident from the left panel of figure \ref{fig:abpp} (for type I seesaw) and figure \ref{fig:abpiss} 
(for inverse seesaw) respectively, where red dots represents allowed parameter 
space and one can find that relatively smaller values for $\alpha$ and $\beta$ 
are allowed for inverse\ seesaw compared to the type-I seesaw scenario. These 
smaller vales of $\alpha$ and $\beta$ for inverse seesaw case actually yields 
larger value for the common factor $\kappa$ (evaluated using equation 
\eqref{kappa34}) appearing 
in the absolute light neutrino masses and hence generates larger value for 
neutrino mass compared to type I case. 
\section{Conclusion}\label{sec:conc}

We have studied two different seesaw scenarios for light Dirac neutrinos namely, 
 type I and inverse seesaw within the framework of $A_4$ flavour symmetry to 
 explain lepton masses and mixing. In both the cases, the $A_4$ 
symmetry is augmented by additional discrete symmetries in order to make sure 
that the correct hierarchy between different terms appearing in the complete 
neutral fermion mass matrix is naturally obtained without making any ad hoc 
assumptions. This is done by generating relatively smaller terms at next to 
leading order compared to the large terms in the seesaw matrix. Since lepton 
number is a global conserved symmetry in both the cases, all the mass matrices 
involved are of Dirac type and hence the $A_4$ triple products contain the 
anti-symmetric component. This anti-symmetric part plays a crucial role in 
generating the correct neutrino phenomenology by explicitly breaking $\mu-\tau$ 
symmetries which give rise to vanishing reactor mixing angle. Since we use the 
$S$ diagonal basis of $A_4$ for Dirac neutrino case, the charged lepton mass 
matrix is also non trivial in our scenarios and hence can contribute to the 
leptonic mixing matrix.

For generic choices of $A_4$ flavon alignments, we find that both the models  
are very predictive in terms of predicting the light neutrino mass spectrum and 
hierarchy, leptonic CP phase as well as the octant of atmospheric mixing angle. 
While both of them predicts normal hierarchical pattern of light neutrino masses 
with the atmospheric mixing angle lying in the lower octant, in agreement with 
the latest global fit neutrino oscillation data, they also predict the leptonic 
Dirac CP phase to lie in specific range -$\pi/3$ to $\pi/3$. While the type I 
seesaw predicts the sum of light neutrino masses to be small, the inverse seesaw 
scenario predicts it to be high and can saturate the cosmological upper bound 
$\sum  m_i  \leq 0.17$ eV. Apart from this, the models also predict 
interesting correlation between neutrino observables like Dirac CP phase, 
atmospheric mixing angle, light neutrino masses so that measuring one can shed 
light on the other. Both the models can also predict the absence of lepton 
number violation and hence can not be tested in ongoing and future neutrinoless 
double beta decay experiments. Also, the inverse seesaw model can naturally 
predict lighter heavy neutrino spectrum compared to type I seesaw and hence can 
have other phenomenological consequences. Such a detailed analysis is left for 
future investigations.

Apart from different predictions for light neutrino parameters,
the two seesaw scenarios discussed here can also be distinguished by observing 
different phenomena they give rise to. Since the light neutrino mass in inverse 
seesaw mechanism is primarily governed by the smallness of the $\mu$ term in 
\eqref{eq:iss1}, the right handed neutrinos can have masses near the TeV scale 
and at the same time can have sizeable Yukawa couplings with the light 
neutrinos, giving rise to interesting possibilities at collider experiments 
\cite{Antusch:2016ejd}. This interesting feature makes it different from 
ordinary type I seesaw, where TeV scale right handed neutrino mass has to be 
compensated by tiny Yukawa couplings. We may also get distinguishable features 
in terms of predictions of these two models, if we also incorporate the quark 
sector mixing \cite{He:2006dk}. For simplicity, we have considered the quark 
sector particles to be singlet under the $A_4$ symmetry and leave a more general 
study including quarks and leptons to future studies.

\section*{Acknowledgements}
The authors would like to thank the organisers of XV Workshop on High Energy  
Physics Phenomenology during 14-23 December, 2017 at Indian Institute of Science 
Education and Research Bhopal, India where part of this work was completed.
\appendix
\section{$A_4$ Multiplication Rules}
\label{appen1}
$A_4$, the symmetry group of a tetrahedron, is a discrete non-abelian group of even permutations of four objects. It has four irreducible representations: three one-dimensional and one three dimensional which are denoted by $\bf{1}, \bf{1'}, \bf{1''}$ and $\bf{3}$ respectively, being consistent with the sum of square of the dimensions $\sum_i n_i^2=12$. We denote a generic permutation $(1,2,3,4) \rightarrow (n_1, n_2, n_3, n_4)$ simply by $(n_1 n_2 n_3 n_4)$. The group $A_4$ can be generated by two basic permutations $S$ and $T$ given by $S = (4321), T=(2314)$. This satisfies 
$$ S^2=T^3 =(ST)^3=1$$
which is called a presentation of the group. Their product rules of the irreducible representations are given as
$$ \bf{1} \otimes \bf{1} = \bf{1}$$
$$ \bf{1'}\otimes \bf{1'} = \bf{1''}$$
$$ \bf{1'} \otimes \bf{1''} = \bf{1} $$
$$ \bf{1''} \otimes \bf{1''} = \bf{1'}$$
$$ \bf{3} \otimes \bf{3} = \bf{1} \otimes \bf{1'} \otimes \bf{1''} \otimes \bf{3}_a \otimes \bf{3}_s $$
where $a$ and $s$ in the subscript corresponds to anti-symmetric and symmetric parts respectively. Denoting two triplets as $(a_1, b_1, c_1)$ and $(a_2, b_2, c_2)$ respectively, their direct product can be decomposed into the direct sum mentioned above. In the $S$ diagonal basis, the products are given as
$$ \bf{1} \backsim a_1a_2+b_1b_2+c_1c_2$$
$$ \bf{1'} \backsim a_1 a_2 + \omega^2 b_1 b_2 + \omega c_1 c_2$$
$$ \bf{1''} \backsim a_1 a_2 + \omega b_1 b_2 + \omega^2 c_1 c_2$$
$$\bf{3}_s \backsim (b_1c_2+c_1b_2, c_1a_2+a_1c_2, a_1b_2+b_1a_2)$$
$$ \bf{3}_a \backsim (b_1c_2-c_1b_2, c_1a_2-a_1c_2, a_1b_2-b_1a_2)$$
In the $T$ diagonal basis on the other hand, they can be written as
$$ \bf{1} \backsim a_1a_2+b_1c_2+c_1b_2$$
$$ \bf{1'} \backsim c_1c_2+a_1b_2+b_1a_2$$
$$ \bf{1''} \backsim b_1b_2+c_1a_2+a_1c_2$$
$$\bf{3}_s \backsim \frac{1}{3}(2a_1a_2-b_1c_2-c_1b_2, 2c_1c_2-a_1b_2-b_1a_2, 2b_1b_2-a_1c_2-c_1a_2)$$
$$ \bf{3}_a \backsim \frac{1}{2}(b_1c_2-c_1b_2, a_1b_2-b_1a_2, c_1a_2-a_1c_2)$$

\end{document}